\documentclass[seceq]{jpsj2}
\usepackage{mathptm,times,graphicx}
\usepackage{amsmath,amssymb,amsthm}
\usepackage{latexsym}
\makeatletter
\newdimen\LENB \newdimen\LENW \newdimen\THI
\newdimen\LENWH \newdimen\LENTOT \newcount\N
\def\vbrknlnele#1#2#3{
  \LENB=#1pt \LENW=#2pt \THI=#3pt
  \LENWH=\LENW \divide\LENWH by 2
  \LENTOT=\LENB \advance\LENTOT by \LENW
  \vbox to \LENTOT{
    \vbox to \LENWH{}
    \nointerlineskip
    \vbox to \LENB{\hbox to \THI{\vrule width \THI height \LENB}}
    \nointerlineskip
    \vbox to \LENWH{}
  }}

\def\vbrknln#1{
  \N=#1
  \vcenter{
    \vbox{
      \loop\ifnum\N>0
        \vbox to 4pt{\vbrknlnele{2}{2}{0.1}}
        \nointerlineskip
        \advance\N by -1
      \repeat
  }}}

\def\vbl#1{\hskip-5pt \vbrknln{#1} \hskip-5pt}

\def\hbrknlnele#1#2#3{
  \LENB=#1pt \LENW=#2pt \THI=#3pt
  \LENTOT=\LENB \advance\LENTOT by \LENW
  \vcenter{
    \vbox to \THI{
      \hbox to \LENTOT{
        \hfil
        \vrule width \LENB height \THI
        \hfil}
  }}}

\def\hblele{\hbrknlnele{2}{2.2}{0.1}}

\def\hblfil{\cleaders\hbox{$ \m@th \mkern1mu \hblele \mkern1mu $}\hfill} 
\makeatother

\title{
Casorati Determinant Form of Dark Soliton Solutions of 
the Discrete Nonlinear Schr\"odinger Equation
}
\author
{Ken-ichi \textsc{Maruno}$^{1}$
\thanks{E-mail: maruno@math.kyushu-u.ac.jp}, 
and Yasuhiro \textsc{Ohta}$^{2}$
}
\inst{$^{1}$Faculty of Mathematics, 
Kyushu University, Hakozaki, Higashi-ku, Fukuoka, 812-8581, Japan\\
$^{2}$
Department of Mathematics, 
Kobe University, Rokko, Kobe 657-8501, Japan}

\abst{
It is shown that the $N$-dark soliton solutions 
of the integrable discrete 
nonlinear Schr\"odinger (IDNLS) equation 
are given in terms of the Casorati determinant. 
The conditions for reduction, complex conjugacy and regularity
for the Casorati determinant solution are also given explicitly.
The relationship between the IDNLS and the relativistic Toda 
lattice is discussed. 
}

\kword{Casorati Determinant, Discrete Nonlinear Schr\"odinger Equation,
Dark Soliton}

\begin{document}
\maketitle

%%%\newcommand{\R}{\mathbb{R}}
%%%\newcommand{\D}{\displaystyle}

%%\numberwithin{equation}{section}

%%%\newtheorem{thm}{Theorem}[section]
%%%\newtheorem{Le}{Lemma}[section]
%%%\newtheorem{defi}{Definition}[section]

%\setlength{\baselineskip}{18pt}
\def \part {\partial}
\def \ba {\begin{array}}
\def \ea {\end{array}}
\def \si {\sigma}
\def \al {\alpha}
\def \la {\lambda}

\section{Introduction}

The nonlinear Schr\"odinger (NLS) equation,
\begin{equation}
{\rm i}\psi_t=\psi_{xx}+\alpha|\psi|^2\psi\,, 
\label{contiNLS}
\end{equation}
is one of most important soliton equations in mathematics and physics. 
The study of discrete analogues of the NLS 
equation has received considerable attention recently 
from both physical and mathematical 
point of view.~\cite{APT,Kevrekidis} 
The integrable discrete nonlinear Schr\"odinger (IDNLS) equation 
is given by 
\begin{equation}
{\rm i}\frac{d\psi_n}{dt}=\psi_{n+1}+\psi_{n-1}+\alpha 
|\psi_n|^2(\psi_{n+1}+\psi_{n-1})\,. \label{ablowitz-ladik}
\end{equation}
The IDNLS equation was originally 
derived by Ablowitz and Ladik using 
the Lax formulation.~\cite{AL,AL2,APT} 
Thus the IDNLS equation is often called the 
Ablowitz-Ladik lattice. 

The IDNLS equation can be bilinearized into
\begin{eqnarray}
&&D_tG_n\cdot F_n=G_{n+1}F_{n-1}-F_{n+1}G_{n-1}\,,
\label{bilinHIROTA1}\\
&&F_n^2+\alpha G_nG_n^*=F_{n+1}F_{n-1}\,,
\label{bilinHIROTA2}
\end{eqnarray}
through the dependent variable transformation 
$\psi_n=({\rm i})^nG_n/F_n$ where $F_n$ is a real function, 
$G_n$ is a complex function and $G_n^*$ is its complex conjugate. 
The Hirota D-operator $D_t$ is defined by
\[
 D_t^nf\cdot g=\left(\partial_t-\partial_{t'}\right)^n
f(t)g(t')|_{t'=t}\,\,. 
\]
With this bilinear forms the IDNLS equation in the case of 
$\alpha=1$ can be 
solved by the Hirota bilinear method, i.e. 
$N$-bright soliton solutions are obtained 
\cite{HirotaBook,Hirota-AL,NakamuraBook,Narita1,Ohta-Pfaffian,Sadakane}.  

It is well known that 
the NLS equation with defocusing parameter ($\alpha<0$)
has dark soliton solutions. 
It is easily confirmed that 
there is a stationary dark soliton solution 
\begin{equation}
\psi_n=\exp(-2{\rm i}t\,{\rm sech}^2(k))
\frac{\tanh(k)}{\sqrt{-\alpha}}\tanh(k n)
\end{equation} 
by the direct substitution of an ansatz of stationary solution 
into eq.(\ref{ablowitz-ladik}).
Understanding various properties of
dark solitons in discrete lattices is important 
from physical point of view 
\cite{Kivshar,Johansson,Kevrekidis2}.  
There are analytical results of 
the IDNLS equation under non-vanishing 
boundary condition using the
perturbative Hirota method by Narita \cite{Narita2}
and inverse scattering theory 
by Vekslerchik and Konotop \cite{Vekslerchik2}. 
However, the detailed anayisis of the solution and 
understanding the role of a lattice space parameter 
in the $N$-dark soliton solutions 
are missing in their results. 
In ref.\citen{Vekslerchik2}, a determinant expression of dark soliton solutions
is given, however for their boundary condition there is no carrier wave
of the dark soliton. Besides this is a Grammian type determinant solution. 
The Casorati determinant solution which is well-known in discrete 
integrable systems is missing.    
%%%The $N$-dark soliton solutions of the IDNLS equation 
%%%have not been studied much until now. 

The relationship between the IDNLS and relativistic Toda lattice (RTL) 
equation was also pointed out by several authors
\cite{Vekslerchik,KMZ97,Sadakane}. 
An interesting open problem was given in conclusion in 
ref.\citen{Sadakane}: 
The 2-component Casorati determinant solution of 
the IDNLS equation and RTL equation in ref.\citen{Sadakane} 
is different from the single-component Casorati determinant 
solution of the RTL equation which was derived in ref.\citen{OKMS}. 
Is there any relationship between the 2-component Casorati determinant 
solution and single-component Casorati determinant solution? 

Our goal in this paper is to construct Casorati determinant 
form of $N$-dark soliton solutions with 
a lattice space parameter, analyze the detail of 
behaviour of 
$N$-dark soliton solutions 
and give an answer of 
the above open problem. 
In this article, we give an explicit formula of 
the dark soliton solutions.
It is known that the dark soliton solutions for NLS are
written in terms of Wronski determinants.
We show that the solution for IDNLS is given in the Casorati
determinant form which is a discrete analogue of the Wronskian.
%%%Letting
%%%\begin{eqnarray}
%%%F_n = \tau_n(n),\quad G_n = \tau_{n-1}(n),
%%%\quad G^*_n = \tau_{n+1}(n),
%%%\end{eqnarray}
%%%we have 
%%%\begin{eqnarray}
%%%&&(iD_t+2)\tau_{n-1}(n)\cdot \tau_{n}(n)\nonumber\\
%%%&&\quad \quad 
%%% =\tau_{n}(n+1)\tau_{n-1}(n-1)
%%%-\tau_{n+1}(n+1)\tau_{n-2}(n-1)\,,\label{al-bilinear1}\\
%%%&&a^2\tau_n(n) \tau_n(n)+
%%%(1-a^2)\tau_{n-1}(n)\tau_{n+1}(n)\nonumber\\
%%%&&\quad \quad \quad \quad =
%%%\tau_{n+1}(n+1)\tau_{n-1}(n-1)\,.\label{al-bilinear2}
%%%\end{eqnarray}
%%%%Introducing new variables
%%%%\begin{equation}
%%%%t_1=-\frac{a}{1-a^2}(it+y)\,,\quad 
%%%t_{-1}=\frac{a}{1-a^2}(-it+y)\,,
%%%%t_{-1}=-\frac{a}{1-a^2}(-it+y)\,,
%%%%\end{equation}
%%%%we can decouple eq.(\ref{al-bilinear1}) into 
%%%%\begin{eqnarray}
%%%%&&(a D_{t_1}+1)\tau_{n-1}(n)\cdot \tau_{n}(n)
%%%%=\tau_{n}(n+1)\tau_{n-1}(n-1)
%%%%\,,\label{al-bilinear1-1}\\
%%%%&&(a D_{t_{-1}}+1)\tau_{n-1}(n)\cdot \tau_{n}(n)
%%%% =\tau_{n+1}(n+1)\tau_{n-2}(n-1)
%%%%\,.\label{al-bilinear1-2}
%%%%\end{eqnarray}

This paper is organized as follows. 
In the next section, we discuss about the gauge transformation
and bilinearization of IDNLS equation.
In \S 3, we give bilinear identities for 
Casorati determinant both for the B\"acklund transformation of
Toda lattice and the discrete 2-dimensional Toda lattice equation. 
In \S 4, we perform a reduction of the Casorati determinant 
which derives the bilinear form for IDNLS
given in \S 2. 
The complex conjugate condition is considered in \S 5. 
In \S 6, we give 
examples of dark soliton solutions. 
We also clarify the
relationship between the IDNLS and the relativistic Toda lattice in 
the final section. 

%%%\section{Bilinear Forms for Dark Soliton Solutions}

\section{Gauge Transformation of IDNLS}

First of all, we should examine the transformation of IDNLS equation
by gauge.
Applying the gauge transformation,
$$
\psi_n=u_nA\exp({\rm i}(\theta n-\omega t)),
$$
eq.~(\ref{ablowitz-ladik}) is transformed as
$$
{\rm i}\frac{du_n}{dt}=-\omega u_n
+(1+\alpha A^2|u_n|^2)
(u_{n+1}\exp({\rm i}\theta)+u_{n-1}\exp(-{\rm i}\theta))\,,
$$
thus we can scale $|\alpha|$ by using the amplitude $A$.
To obtain the dark soliton solutions, we set 
$\alpha=-(a^2-1)/a^2$, $(a > 1)$.
We also scale $t$ by $t\to a^2t$,
then the IDNLS eq.~(\ref{ablowitz-ladik}) can be bilinearized into
\begin{eqnarray}
 &&({\rm i}D_t+\omega)G_n\cdot F_n=
G_{n+1}F_{n-1}\exp({\rm i}\theta)
+G_{n-1}F_{n+1}\exp(-{\rm i}\theta)\,,
\label{bilin-gauge1}\\
&&a^2F_n^2-(a^2-1)G_nG_n^*= F_{n+1}F_{n-1}\,,
\label{bilin-gauge2}
\end{eqnarray}
through the dependent variable transformation,
$$
\psi_n=\frac{G_n}{F_n}\exp({\rm i}(\theta n-\omega t)),
$$
where $F_n$ is a real function, 
$G_n$ is a complex function and $G_n^*$ is its complex conjugate. 

The above gauge factor $\exp({\rm i}(\theta n-\omega t))$ is
nothing but the carrier wave of the dark soliton, and essentially
$u_n=G_n/F_n$ gives the envelope of the soliton solution.
In the context of NLS, in order to construct physical solutions,
it is more convenient to consider the above bilinear form than
the one in eqs.~(\ref{bilinHIROTA1}) and (\ref{bilinHIROTA2}).
Using the bilinear forms, eqs.~(\ref{bilin-gauge1})
and (\ref{bilin-gauge2}),
we present the Casorati determinant form of 
dark soliton solutions of the IDNLS equation. 
Instead of applying the Hirota bilinear method directly to
eqs.~(\ref{bilin-gauge1}) and (\ref{bilin-gauge2}),
we rather start from the algebraic identities for Casorati
determinant and derive the bilinear form of IDNLS equation
by using the reduction technique.

\section{Bilinear Identities for Casorati Determinant}

\subsection{B\"acklund transformation of Toda lattice}
We have the bilinear forms for B\"acklund Transformation (BT) 
of Toda lattice (TL) equation,
\begin{eqnarray}
&&(aD_x-1)\tau_{n+1}(k+1,l)\cdot \tau_{n}(k,l)
+\tau_{n}(k+1,l)\tau_{n+1}(k,l)=0\,,\label{bt-bilinear-1}\\
&&(bD_y-1)\tau_{n-1}(k,l+1)\cdot 
\tau_n(k,l)+\tau_n(k,l+1)\tau_{n-1}(k,l)=0\,,
\label{bt-bilinear-2}
%%%(aD_{t_{-1}}+1)\tau_{n-1}(k,l+1)\cdot 
%%%\tau_n(k,l)=\tau_n(k,l-1)\tau_{n-1}(k,l+2)\,,
\end{eqnarray}
where $a$ and $b$ are constants.  
\par
Consider the following Casorati determinant solution,
\begin{eqnarray}
 \tau_n(k,l) = \left|\begin{array}{cccc}
  \varphi_1^{(n)}(k,l) &\varphi_1^{(n+1)}(k,l) &\cdots
   &\varphi_1^{(n+N-1)}(k,l) \cr
  \varphi_2^{(n)}(k,l) &\varphi_2^{(n+1)}(k,l) &\cdots
   &\varphi_2^{(n+N-1)}(k,l) \cr
  \vdots               &\vdots                 &
   &\vdots                 \cr
  \varphi_N^{(n)}(k,l) &\varphi_N^{(n+1)}(k,l) &\cdots
   &\varphi_N^{(n+N-1)}(k,l) \cr
\end{array}
 \right|,\label{determinant-bt}
\end{eqnarray}
where $\varphi_i^{(n)}$'s are arbitrary functions of two continuous
independent variables, $x$ and $y$, and two discrete ones, $k$ and $l$, which
satisfy the dispersion relations,
\begin{eqnarray}
&&\partial_x\varphi_i^{(n)}(k,l) = \varphi_i^{(n+1)}(k,l),
\label{t1-dispersion}\\
&&\partial_y\varphi_i^{(n)}(k,l) = \varphi_i^{(n-1)}(k,l),
\label{t-1-dispersion}\\
&&\Delta_k\varphi_i^{(n)}(k,l) 
= \varphi_i^{(n+1)}(k,l),\label{k-linear}\\
&&
 \Delta_l\varphi_i^{(n)}(k,l) = \varphi_i^{(n-1)}(k,l).\label{l-linear}
\end{eqnarray}
Here $\Delta_k$ and $\Delta_l$ are 
the backward difference operators with difference intervals
$a$ and $b$, defined by
\begin{eqnarray}
&& \Delta_k f(k,l) = {f(k,l)-f(k-1,l) \over a},\label{def-k-bt}\\
&& \Delta_l f(k,l) = {f(k,l)-f(k,l-1) \over b}.\label{def-l-bt}
\end{eqnarray}
For simplicity we introduce a convenient notation,
\begin{eqnarray}
|{n_1}_{{}_{\scriptstyle k_1,l_1}},
  {n_2}_{{}_{\scriptstyle k_2,l_2}},
  \cdots,
  {n_N}_{{}_{\scriptstyle k_N,l_N}}
|
 = \left|\begin{matrix}
  \varphi_1^{(n_1)}(k_1,l_1) &\varphi_1^{(n_2)}(k_2,l_2) &\cdots
   &\varphi_1^{(n_N)}(k_N,l_N) \cr
  \varphi_2^{(n_1)}(k_1,l_1) &\varphi_2^{(n_2)}(k_2,l_2) &\cdots
   &\varphi_2^{(n_N)}(k_N,l_N) \cr
  \vdots                     &\vdots                     &
   &\vdots                     \cr
  \varphi_N^{(n_1)}(k_1,l_1) &\varphi_N^{(n_2)}(k_2,l_2) &\cdots
   &\varphi_N^{(n_N)}(k_N,l_N) \cr
\end{matrix}
 \right|.
\end{eqnarray}
In this notation, the solution for BT of TL, $\tau_n(k,l)$ 
in eq.~(\ref{determinant-bt}),
is rewritten as
\begin{eqnarray}
 \tau_n(k,l) = |n_{k,l},n+1_{k,l},\cdots,n+N-1_{k,l}|,
\end{eqnarray}
or suppressing the index $k$ and $l$, we simply write as
$$
 \tau_n(k,l) = |n,n+1,\cdots,n+N-1|.
$$
\par
 We show that the above $\tau_n(k,l)$ actually satisfies 
the bilinear eqs.~(\ref{bt-bilinear-1}) 
and (\ref{bt-bilinear-2}) by using the Laplace expansion
technique\cite{Freeman-Nimmo1,Freeman-Nimmo2}. 
At first we investigate the difference formula for $\tau$.
{}From eq.~(\ref{k-linear}) we have
$$
 a\varphi_i^{(n+1)}(k+1,l) = \varphi_i^{(n)}(k+1,l) - \varphi_i^{(n)}(k,l).
$$
Noticing this relation, we get
\begin{eqnarray}
  \tau_n(k+1,l) &=& 
|n_{k+1,l},n+1_{k+1,l},n+2_{k+1,l},\cdots,n+N-1_{k+1,l}| \cr
              &=& 
|n_{k,l},n+1_{k+1,l},n+2_{k+1,l},\cdots,n+N-1_{k+1,l}| \cr
\noalign{\hbox{where we have subtracted the 2nd column multiplied by $a$
from the 1st column, }}
              &=& 
|n_{k,l},n+1_{k,l},n+2_{k+1,l},\cdots,n+N-1_{k+1,l}| \cr
\noalign{\hbox{where we have subtracted the 3rd column multiplied by $a$
from the 2nd column, }}
              &&\cdots \cr
              &=& 
|n_{k,l},n+1_{k,l},\cdots,n+N-2_{k,l},n+N-1_{k+1,l}|,
\end{eqnarray}
that is,
\begin{equation}
 \tau_n(k+1,l) =
 |n_{k,l},n+1_{k,l},\cdots,n+N-2_{k,l},n+N-1_{k+1,l}|.
\label{k-shift-tau1-bt}
\end{equation}
Moreover in eq.~(\ref{k-shift-tau1-bt}), 
multiplying the $N$-th column by $a$ and adding
the $(N-1)$-th column to the $N$-th column, we get
\begin{equation}
 a\tau_n(k+1,l) =
 |n_{k,l},n+1_{k,l},\cdots,n+N-2_{k,l},n+N-2_{k+1,l}|.
\label{a-k-shift-tau1-bt}
\end{equation}
Differentiating eq.(\ref{a-k-shift-tau1-bt}) with 
$x$ and using eq.(\ref{t1-dispersion}) we obtain
\begin{eqnarray}
&& a\partial_x\tau_n(k+1,l) =
 |n_{k,l},n+1_{k,l},\cdots,n+N-3_{k,l},n+N-1_{k,l},n+N-2_{k+1,l}|
\nonumber\\
&& \quad \quad \quad \quad \quad \quad \quad 
+ |n_{k,l},n+1_{k,l},\cdots,n+N-3_{k,l},n+N-2_{k,l},n+N-1_{k+1,l}|
\nonumber\\
&& \quad =
 |n_{k,l},n+1_{k,l},\cdots,n+N-3_{k,l},n+N-1_{k,l},n+N-2_{k+1,l}|
+\tau_n(k+1,l)\,.
\end{eqnarray}
We have also 
\begin{equation}
\partial_x\tau_n(k,l)=
 |n_{k,l},n+1_{k,l},\cdots,n+N-2_{k,l},n+N_{k,l}|\,.
\end{equation}
In short, we write
\begin{equation}
 \tau_n(k+1,l) = |n,n+1,\cdots,n+N-2,n+N-1_{k+1}|\,,
\label{k-shift-tau-bt}
\end{equation}
\begin{equation}
 a\tau_n(k+1,l) = |n,n+1,\cdots,n+N-2,n+N-2_{k+1}|\,,
\label{k-shift-tau-bt-2}
\end{equation}
\begin{equation}
 (a\partial_x-1)\tau_n(k+1,l)
=|n,n+1,\cdots,n+N-3,n+N-1,n+N-2_{k+1}|\,,
\label{k-shift-tau-bt-3}
\end{equation}
\begin{equation}
 \partial_x\tau_n(k,l)
=|n,n+1,\cdots,n+N-2,n+N|\,.
\label{k-shift-tau-bt-4}
\end{equation}

Let us introduce an identity for $2N\times 2N$ determinant,
$$
\left|\begin{array}{cccccccccccccc}
n+1 &\cdots &n+N-2 & \vbl4 &n+N &\vbl4
&n+N-1_{k+1} &\vbl4 &     &    &\hbox{\O} &  &\vbl4 &n+N-1 \cr
\multispan{14}\hblfil \cr
&\hbox{\O} &        &\vbl4 &n+N &\vbl4 &n+N-1_{k+1} &\vbl4
&n &n+1 &\cdots &n+N-2 &\vbl4 &n+N-1\cr
\end{array}
 \right| = 0.
$$
Applying the Laplace expansion to the left-hand side, we obtain the
algebraic bilinear identity for determinants,
\begin{eqnarray}
 &|n+1,\cdots,n+N-2,n+N,n+N-1_{k+1}|\times
     |n,n+1,\cdots,n+N-2,n+N-1| \cr
  - &|n+1,\cdots,n+N-2,n+N-1,n+N-1_{k+1}|\times
     |n,n+1,\cdots,n+N-2,n+N| \cr
  - &|n+1,\cdots,n+N-2,n+N,n+N-1|\times
     |n,n+1,\cdots,n+N-2,n+N-1_{k+1}|
= 0,&
\end{eqnarray}
which is rewritten by using 
eqs.~(\ref{k-shift-tau-bt})-(\ref{k-shift-tau-bt-4}), into
the differential bilinear equation,
$$
   (a\partial_x-1)\tau_{n+1}(k+1,l)\times\tau_n(k,l)
 - a\tau_{n+1}(k+1,l)\partial_x\tau_n(k,l)
 + \tau_{n+1}(k,l)\tau_n(k+1,l) = 0.
$$
This equation is equal to eq.~(\ref{bt-bilinear-1}). 

Similarly we obtain
$$
(b\partial_y-1)\tau_{n-1}(k,l+1)\times\tau_n(k,l)
 - b\tau_{n-1}(k,l+1)\partial_y\tau_n(k,l)
 + \tau_{n-1}(k,l)\tau_n(k,l+1) = 0,
$$ 
because the role of $y$, $l$ and $b$ is parallel to that of
$x$, $k$ and $a$ except that the ordering of index $n$ is reversed.
The above equation is equal to eq.~(\ref{bt-bilinear-2}). 
Hence we have proved 
that the Casorati determinant $\tau_n(k,l)$ 
in eq.~(\ref{determinant-bt}) gives the solution for the BT of TL
equation.

\subsection{Discrete 2-dimensional Toda lattice}
Using the Casoratian technique, we show that the bilinear form of 
discrete 2-dimensional Toda lattice (D2DTL) equation,
\begin{eqnarray}
&& \tau_n(k+1,l+1)\tau_n(k,l) - \tau_n(k+1,l)\tau_n(k,l+1)\cr
&&\quad =  
ab( \tau_n(k+1,l+1)\tau_n(k,l) - 
\tau_{n+1}(k+1,l)\tau_{n-1}(k,l+1) ),\label{d2dTL}
\end{eqnarray}
is also
satisfied by the Casorati determinant.  

Let us further examine the difference formula for the Casorati
determinant $\tau_n$ in eq.~(\ref{determinant-bt}).
Similarly to eqs.~(\ref{k-shift-tau-bt}) and (\ref{k-shift-tau-bt-2}),
we have
\begin{equation}
 \tau_n(k,l+1) = |n_{l+1},n+1,\cdots,n+N-2,n+N-1|,
\label{l-shift-tau}
\end{equation}
\begin{equation}
 b\tau_n(k,l+1) = |n+1_{l+1},n+1,\cdots,n+N-2,n+N-1|.
\label{l-shift-tau2}
\end{equation}
Thus $\tau_n(k+1,l+1)$ is given by
\begin{eqnarray}
\tau_n(k+1,l+1) =
 |n_{k+1,l+1},n+1_{k+1},\cdots,n+N-2_{k+1},n+N-1_{k+1}|.
\label{k-l-shift-tau}
\end{eqnarray}
In the following we show that the shifts of index $k$ is condensed into
only the right-most column of the determinant.
{}From eqs.~(\ref{k-linear}) and (\ref{l-linear}), 
$\varphi_i^{(n)}$
satisfies
\begin{equation}
 \Delta_l\Delta_k\varphi_i^{(n)} = \varphi_i^{(n)},
\end{equation}
that is,
\begin{equation}
  \varphi_i^{(n)}(k,l) - \varphi_i^{(n)}(k-1,l) 
- \varphi_i^{(n)}(k,l-1)
  + \varphi_i^{(n)}(k-1,l-1) = ab\varphi_i^{(n)}(k,l),
\end{equation}
which is rewritten as
\begin{eqnarray}    
(1-ab)\varphi_i^{(n)}(k+1,l+1)
 &=& \varphi_i^{(n)}(k,l+1) 
+ \varphi_i^{(n)}(k+1,l) - \varphi_i^{(n)}(k,l) \cr
 &=& \varphi_i^{(n)}(k,l+1) + a\varphi_i^{(n+1)}(k+1,l).
\label{1-ab}
\end{eqnarray}
Therefore multiplying the both hand sides of 
eq.~(\ref{k-l-shift-tau}) by $(1-ab)$ and
rewriting the 1st column of the determinant by the use of 
eq.~(\ref{1-ab}),
we obtain
\begin{eqnarray}
&&  (1-ab)\tau_n(k+1,l+1)\cr
  &=& |n_{k,l+1},n+1_{k+1},\cdots,n+N-2_{k+1},n+N-1_{k+1}| \cr
  &+& a|n+1_{k+1,l},n+1_{k+1},\cdots,n+N-2_{k+1},n+N-1_{k+1}|.
\end{eqnarray}
The second term in r.h.s. vanishes because its 1st and 2nd columns are
the same.  So we get
\begin{equation}
  (1-ab)\tau_n(k+1,l+1) = |n_{l+1},n+1_{k+1},\cdots,n+N-2_{k+1},n+N-1_{k+1}|.
\end{equation}
In the above determinant, by subtracting the $(j+1)$-th 
column multiplied by
$a$ from the $j$-th column for $j=2,3,\cdots,N-1$, 
$\tau_n(k+1,l+1)$
is given by
\begin{equation}
(1-ab)\tau_n(k+1,l+1) = |n_{l+1},n+1,\cdots,n+N-2,n+N-1_{k+1}|.
\label{k-l-shift-tau-ab}
\end{equation}
As is shown, even if the two variables $k$ and $l$ are shifted, $\tau_n$
is also expressed in the determinant form whose columns are
almost unchanged and only edges are varied.
Thus we can use the Laplace expansion technique.  

We consider an identity for $2N\times 2N$ determinant,
$$
 \left|\begin{array}{cccccccccccccc}
  n_{l+1} &\vbl4 &n+1 &\cdots    &n+N-2 &n+N-1_{k+1} &\vbl4 &n &\vbl4
   &    &\hbox{\O} &      &\vbl4 &n+N-1 \cr
 \multispan{14}\hblfil \cr
  n_{l+1} &\vbl4 &    &\hbox{\O} &      &            &\vbl4 &n &\vbl4
   &n+1 &\cdots    &n+N-2 &\vbl4 &n+N-1 \cr
\end{array}
 \right| = 0.
$$
By the Laplace expansion, we have
\begin{eqnarray}
    &&|n_{l+1},n+1,\cdots,n+N-2,n+N-1_{k+1}|\times
|n,n+1,\cdots,n+N-2,n+N-1| \cr
  - &&|n,n+1,\cdots,n+N-2,n+N-1_{k+1}|\times
|n_{l+1},n+1,\cdots,n+N-2,n+N-1| \cr
  + &&|n+1,\cdots,n+N-2,n+N-1,n+N-1_{k+1}|\times
|n_{l+1},n,n+1,\cdots,n+N-2| = 0.
\label{laplace}
\end{eqnarray}
Using eqs.~(\ref{k-shift-tau-bt}), (\ref{k-shift-tau-bt-2}),
(\ref{l-shift-tau}), (\ref{l-shift-tau2}) 
and (\ref{k-l-shift-tau-ab}),
we obtain from eq.~(\ref{laplace}),
\begin{eqnarray}
&& (1-ab)\tau_n(k+1,l+1)\tau_n(k,l) - 
\tau_n(k+1,l)\tau_n(k,l+1)\cr
&& + a\tau_{n+1}(k+1,l)b\tau_{n-1}(k,l+1) = 0,
\end{eqnarray}
which recovers eq.~(\ref{d2dTL}).  
This completes the proof that the Casorati
determinant is indeed the solution for D2DTL equation.  \par
\ \par

For example, we can take $\varphi_i^{(n)}$ as
\begin{equation}
  \varphi_i^{(n)}(k,l)
 = p_i^n(1-ap_i)^{-k}
\left(1-b{1 \over p_i}\right)^{-l}\exp(\xi_i)
 + q_i^n(1-aq_i)^{-k}
\left(1-b{1 \over q_i}\right)^{-l}\exp(\eta_i),
\label{phi}
\end{equation}
$$
 \xi_i = p_ix +{1 \over p_i}y+ \xi_{i0},
\quad 
 \eta_i = q_ix +{1 \over q_i}y+ \eta_{i0},
$$
where $p_i$, $q_i$ and $\xi_{i0}$, $\eta_{i0}$ are arbitrary constants
which correspond to the wave numbers and phase parameters of solitons,
respectively.  It can easily be seen that $\varphi_i^{(n)}$ 
in eq.~(\ref{phi})
actually satisfies the dispersion relations,
eqs.~(\ref{t1-dispersion})-(\ref{l-linear}).  \par

\section{Reduction to IDNLS}
In this section, we give the reduction technique in order to derive
the IDNLS equation from the system of BT of TL and D2DTL.
In eq.~(\ref{phi}), we apply a condition for the wave numbers,
\begin{equation}
 q_i = -p_i{1-b{1 \over p_i} \over 1-ap_i}.\label{reduction-1}
\end{equation}
Then we get
\begin{equation}
 p_i = -q_i{1-b{1 \over q_i} \over 1-aq_i},
\end{equation}
and thus
\begin{equation}
 p_i^2{1-b{1 \over p_i} \over 1-ap_i}
  = q_i^2{1-b{1 \over q_i} \over 1-aq_i}.\label{reduction-1-pq}
\end{equation}
On this condition, $\varphi_i^{(n)}$ in eq.~(\ref{phi}) satisfies
\begin{equation}
\varphi_i^{(n+2)}(k+1,l-1) = 
p_i^2{1-b{1 \over p_i} \over 1-ap_i}
\varphi_i^{(n)}(k,l).
\end{equation}
Hence we obtain
\begin{equation}
 \tau_{n+2}(k+1,l-1) =
 \left(\prod_{i=1}^N p_i^2{1-b{1 \over p_i} \over 1-ap_i}\right)
 \tau_n(k,l).\label{red-tau}
\end{equation}
By using eq.~(\ref{red-tau}), the bilinear forms for BT of TL
and D2DTL, eqs.~(\ref{bt-bilinear-1}), (\ref{bt-bilinear-2})
and (\ref{d2dTL}), are rewritten as
\begin{eqnarray}
&& (aD_x-1)\tau_{n+1}(k+1,l)\cdot \tau_n(k,l)
+\tau_{n}(k+1,l)\tau_{n+1}(k,l)=0\,,
\label{bt-tl-bilinear-final-1}\\
&& (bD_y-1)\tau_{n+1}(k+1,l)\cdot 
\tau_n(k,l)+\tau_{n+2}(k+1,l)\tau_{n-1}(k,l)=0\,,
\label{bt-tl-bilinear-final-2}\\
&&   
\tau_{n+1}(k+1,l)\tau_{n-1}(k-1,l)-\tau_{n+1}(k,l)\tau_{n-1}(k,l)
\nonumber\\
&&
\qquad =ab(\tau_{n+1}(k+1,l)\tau_{n-1}(k-1,l)-\tau_n(k,l)\tau_n(k,l)).
\label{dtl-bilinear-final}
\end{eqnarray}
Here we may drop $l$-dependence, thus for simplicity we take $l=0$
hereafter.  
Equation (\ref{dtl-bilinear-final}) 
is nothing but the bilinear form of the discrete
one-dimensional TL equation.  

Let us substitute
\begin{equation}
x={\rm i}act,\qquad y={\rm i}bdt,
\end{equation}
where $c$ and $d$ are constants.
Then we have
\begin{equation}
-{\rm i}\partial_t=ac\partial_x+bd\partial_y,
\end{equation}
and from the bilinear eqs.~(\ref{bt-tl-bilinear-final-1}) and
(\ref{bt-tl-bilinear-final-2}), we get
\begin{equation}
(-{\rm i}D_t-c-d)\tau_{n+1}(k+1)\cdot \tau_n(k)
+c\tau_{n}(k+1)\tau_{n+1}(k)+d\tau_{n+2}(k+1)\tau_{n-1}(k)=0\,.
\label{bt-tl-bilinear-final-1-2}
\end{equation}
The Casorati determinant solution of 
eqs.~(\ref{bt-tl-bilinear-final-1-2}) and
(\ref{dtl-bilinear-final}) (with $l=0$) is given by
$$
 \tau_n(k) = \left|\begin{matrix}
  \varphi_1^{(n)}(k) 
&\varphi_1^{(n+1)}(k) &\cdots 
&\varphi_1^{(n+N-1)}(k) \cr
  \varphi_2^{(n)}(k) 
&\varphi_2^{(n+1)}(k) &\cdots 
&\varphi_2^{(n+N-1)}(k) \cr
  \vdots 
&\vdots  & &\vdots \cr
\varphi_N^{(n)}(k) &\varphi_N^{(n+1)}(k) &\cdots 
&\varphi_N^{(n+N-1)}(k) \cr
\end{matrix}
 \right|,
$$
$$
 \varphi_i^{(n)}(k)
 = p_i^n(1-ap_i)^{-k}\exp(\xi_i)
 + q_i^n(1-aq_i)^{-k}\exp(\eta_i),
$$
$$
 \xi_i = {\rm i} \left(acp_i+{bd \over p_i}\right)t + \xi_{i0},
\qquad
 \eta_i = {\rm i} \left(acq_i+{bd \over q_i}\right)t + \eta_{i0}.
$$
Here $p_{i}$ and $q_{i}$ are related by eq.~(\ref{reduction-1}).

By defining
\begin{equation}
f_{n}=\tau_{n}(0),\qquad
g_{n}=\tau_{n+1}(1),\qquad
h_{n}=\tau_{n-1}(-1),
\label{f-tau}
\end{equation}
from eqs.~(\ref{dtl-bilinear-final}) and
(\ref{bt-tl-bilinear-final-1-2}),
we obtain the bilinear form of IDNLS equation,
\begin{eqnarray}
&&(-{\rm i}D_t-c-d)g_n\cdot f_n+dg_{n+1}f_{n-1}+cg_{n-1}f_{n+1}=0\,,
\label{bilin-1}\\
&&({\rm i}D_t-c-d)h_n\cdot f_n+ch_{n+1}f_{n-1}+dh_{n-1}f_{n+1}=0\,,
\label{bilin-2}\\
&&f_{n+1}f_{n-1}-f_nf_n=(ab-1)(f_nf_n-g_nh_n)\,.
\label{bilin-3}
\end{eqnarray}
By using the dependent variable transformation,
$$
u_n={g_n \over f_n},\qquad
v_n={h_n \over f_n},
$$
the above bilinear equations are transformed into
\begin{eqnarray}
&&{\rm i}{du_n \over dt}
=-(c+d)u_n+(ab-(ab-1)u_nv_n)(du_{n+1}+cu_{n-1})\,,\\
&&-{\rm i}{dv_n \over dt}
=-(c+d)v_n+(ab-(ab-1)u_nv_n)(cv_{n+1}+dv_{n-1})\,.
\end{eqnarray}
\section{Complex Conjugate Condition}
In order to take $v_n$ to be the complex conjugate of $u_n$,
we need to restrict $f_n$ as real and $h_n$ as the complex conjugate
of $g_n$ up to gauge freedom.
Moreover to take $u_n$ to be regular (i.e., $u_n$ not to diverge
for real $t$), we need $f_n\ne0$ for real $t$.
In this section, we give the conditions for the complex conjugacy
and regularity for the Casorati determinant solution.

Firstly we take
$$
b=a\,,
$$
for simplicity, and next we take
$$
a>1\,.
$$
Then the conditions for complex conjugate are given as follows:
\begin{eqnarray}
&&p_i=a+\sqrt{a^2-1}r_i,\qquad |r_i|=1\,,
\label{condcc1}\\
&&\exp(\eta_{i0})=\left(\prod_{{k=1 \atop k\ne i}}^N
{p_i-q_k \over q_i-q_k}\right)\exp(-\xi_{i0}^{*})\,,
\label{condcc2}\\
&&d=c^{*}\,,
\label{condcc3}
\end{eqnarray}
where $r_i$ is a complex parameter of absolute value $1$, and
${}^{*}$ means the complex conjugate.
On the condition (\ref{condcc1}), eq.~(\ref{reduction-1}) turns to be
$$
q_i={1 \over p_i^{*}}\,,
$$
and eqs.~(\ref{condcc2}) and (\ref{condcc3}) gives
$$
\exp(\eta_{i})=\left(\prod_{{k=1 \atop k\ne i}}^N
{p_i-q_k \over q_i-q_k}\right)\exp(-\xi_{i}^{*})\,,
$$
for real $t$.

Now we summarize the final result of the Casorati determinant
solution for IDNLS equation.
The bilinear form of IDNLS equation is given as
\begin{eqnarray}
&&(-{\rm i}D_t-c-c^{*})g_n\cdot f_n+c^{*}g_{n+1}f_{n-1}+cg_{n-1}f_{n+1}=0\,,
\label{bilin-1-b=a}\\
&&({\rm i}D_t-c-c^{*})h_n\cdot f_n+ch_{n+1}f_{n-1}+c^{*}h_{n-1}f_{n+1}=0\,,
\label{bilin-2-b=a}\\
&&f_{n+1}f_{n-1}-f_nf_n=(a^2-1)(f_nf_n-g_nh_n)\,,
\label{bilin-3-b=a}
\end{eqnarray}
where $a>1$ and t is real.
The Casorati determinant solution for the above bilinear equations
is given as
\begin{eqnarray}
&&f_n = \left|\begin{matrix}
  \varphi_1^{(n)}(0) 
&\varphi_1^{(n+1)}(0) &\cdots 
&\varphi_1^{(n+N-1)}(0) \cr
  \varphi_2^{(n)}(0) 
&\varphi_2^{(n+1)}(0) &\cdots 
&\varphi_2^{(n+N-1)}(0) \cr
  \vdots 
&\vdots  & &\vdots \cr
\varphi_N^{(n)}(0) &\varphi_N^{(n+1)}(0) &\cdots 
&\varphi_N^{(n+N-1)}(0) \cr\end{matrix}
 \right|,\\
&&g_n = \left|\begin{matrix}
  \varphi_1^{(n+1)}(1) 
&\varphi_1^{(n+2)}(1) &\cdots 
&\varphi_1^{(n+N)}(1) \cr
  \varphi_2^{(n+1)}(1) 
&\varphi_2^{(n+2)}(1) &\cdots 
&\varphi_2^{(n+N)}(1) \cr
  \vdots 
&\vdots  & &\vdots \cr
\varphi_N^{(n+1)}(1) &\varphi_N^{(n+2)}(1) &\cdots 
&\varphi_N^{(n+N)}(1) \cr\end{matrix}
 \right|,\\
&&h_n = \left|\begin{matrix}
  \varphi_1^{(n-1)}(-1) 
&\varphi_1^{(n)}(-1) &\cdots 
&\varphi_1^{(n+N-2)}(-1) \cr
  \varphi_2^{(n-1)}(-1) 
&\varphi_2^{(n)}(-1) &\cdots 
&\varphi_2^{(n+N-2)}(-1) \cr
  \vdots 
&\vdots  & &\vdots \cr
\varphi_N^{(n-1)}(-1) &\varphi_N^{(n)}(-1) &\cdots 
&\varphi_N^{(n+N-2)}(-1) \cr\end{matrix}
 \right|,
\end{eqnarray}
where
$$
 \varphi_i^{(n)}(k)
 = p_i^n(-\sqrt{a^2-1}p_i^{*}r_i)^{-k}\exp(\xi_i)
 + \left({1 \over p_i^{*}}\right)^n
   \left({\sqrt{a^2-1} \over p_i^{*}r_i}\right)^{-k}
   \left(\prod_{{k=1 \atop k\ne i}}^N
   {p_k^{*}p_i-1 \over p_k^{*}/p_i^{*}-1}\right)\exp(-\xi_{i}^{*}),
$$
$$
 \xi_i = {\rm i} a\left(cp_i+{c^{*} \over p_i}\right)t + \xi_{i0},\qquad
 p_i=a+\sqrt{a^2-1}r_i,\qquad |r_i|=1\,,
$$
where $r_i$ is a complex parameter of absolute value $1$ and
$\xi_{i0}$ is an arbitrary complex parameter.
This Casorati determinant solution gives the $N$-dark soliton
solution for IDNLS equation.
$r_i$ and $\xi_{i0}$ parametrize the wave number and phase constant of
$i$-th soliton, respectively.

After a straightforward and tedious calculation
of expanding the determinants,
$f_n$, $g_n$ and $h_n$ can be expressed in an explicit way
as follows:
\begin{equation}
f_n=F_n{\cal G}\,,\qquad
g_n=G_n{\cal G}{r_1r_2\cdots r_N \over (\sqrt{a^2-1})^N}\,,\qquad
h_n=H_n{\cal G}{(\sqrt{a^2-1})^N \over r_1r_2\cdots r_N}\,,
\label{gaugetrans}
\end{equation}
where ${\cal G}$ is the gauge factor defined by
$$
{\cal G}=
\prod_{N\ge i>j\ge1}(q_i-q_j)\prod_{i=1}^Nq_i^n\exp(\eta_i)\,,
$$
and $F_n$, $G_n$ and $H_n$ are given as
\begin{eqnarray}
&&
F_n=\sum_{M=0}^N\ \sum_{1\le i_1<i_2<\cdots<i_M\le N}
\left|\prod_{1\le\mu<\nu\le M}
{p_{i_\mu}-p_{i_\nu} \over p_{i_\mu}p_{i_\nu}^{*}-1}\right|^2
\prod_{\nu=1}^M
|p_{i_\nu}|^{2n}{\rm e}^{\xi_{i_\nu}+\xi_{i_\nu}^{*}}\,,
\label{expF}\\
&&
G_n=\sum_{M=0}^N\ \sum_{1\le i_1<i_2<\cdots<i_M\le N}
\left|\prod_{1\le\mu<\nu\le M}
{p_{i_\mu}-p_{i_\nu} \over p_{i_\mu}p_{i_\nu}^{*}-1}\right|^2
\prod_{\nu=1}^M\left(-{p_{i_\nu} \over p_{i_\nu}^{*}r_{i_\nu}^2}\right)
|p_{i_\nu}|^{2n}{\rm e}^{\xi_{i_\nu}+\xi_{i_\nu}^{*}}\,,
\label{expG}\\
&&
H_n=\sum_{M=0}^N\ \sum_{1\le i_1<i_2<\cdots<i_M\le N}
\left|\prod_{1\le\mu<\nu\le M}
{p_{i_\mu}-p_{i_\nu} \over p_{i_\mu}p_{i_\nu}^{*}-1}\right|^2
\prod_{\nu=1}^M\left(-{p_{i_\nu}^{*}r_{i_\nu}^2 \over p_{i_\nu}}\right)
|p_{i_\nu}|^{2n}{\rm e}^{\xi_{i_\nu}+\xi_{i_\nu}^{*}}\,.
\label{expH}
\end{eqnarray}
A proof of the above expression is given in appendix B\ref{appproof}.
Now it is clear that $F_n$ is real and $H_n=G_n^{*}$,
thus $v_n$ is the complex conjugate of $u_n$
up to multiplication factor.
Moreover since $F_n>1$, the regularity of $u_n$ is also satisfied.

{}From the bilinear eqs.~(\ref{bilin-1-b=a})-(\ref{bilin-3-b=a}),
and the gauge transformation (\ref{gaugetrans}),
the above $F_n$ and $G_n$ satisfy the bilinear equations,
\begin{eqnarray}
&&(-{\rm i}D_t-c-c^{*})G_n\cdot F_n
+c^{*}G_{n+1}F_{n-1}+cG_{n-1}F_{n+1}=0\,,\\
&&F_{n+1}F_{n-1}-F_nF_n=(a^2-1)(F_nF_n-G_nG_n^{*})\,.
\end{eqnarray}
Let us take
$$
c=\exp(-{\rm i}\theta),
$$
where $\theta$ is a real parameter.
By the variable transformation,
$$
\psi_n={G_n \over F_n}\exp({\rm i}(\theta n-\omega t))\,,
%%%U_n={G_n \over F_n}\,,
$$
that is,
$$
\psi_n=u_n{(\sqrt{a^2-1})^N \over r_1r_2\cdots r_N}
\exp({\rm i}(\theta n-\omega t))\,,
%%%U_n=u_n{(\sqrt{a^2-1})^N \over r_1r_2\cdots r_N}\,,
$$
where
$$
\omega=\exp({\rm i}\theta)+\exp(-{\rm i}\theta)=2\cos\theta\,,
$$
we finally obtain the IDNLS equation,
\begin{equation}
{{\rm i} \over a^2}
\frac{d\psi_n}{dt}=\psi_{n+1}+\psi_{n-1}-\frac{a^2-1}{a^2} 
|\psi_n|^2(\psi_{n+1}+\psi_{n-1})\,.
\label{final-IDNLS}
%%%{\rm i}{dU_n \over dt}
%%%=U_{n+1}-2U_n+U_{n-1}+(a^2-1)(1-|U_n|^2)(U_{n+1}+U_{n-1})\,.
\end{equation}
Thus we have completed the proof that the
$N$-dark soliton solutions for the IDNLS equation are given 
in terms of the Casorati determinant. 
Here we remark that a parameter $a$ is related to 
a lattice space parameter $\delta=1/a\, (<1)$. 
Using a lattice parameter $\delta$, we can rewrite 
the IDNLS equation (\ref{final-IDNLS})
into 
\begin{equation}
{\rm i}
\frac{d\psi_n}{dt}=\frac{\psi_{n+1}+\psi_{n-1}}{\delta^2}
-\frac{1-\delta^2}{\delta^2} 
|\psi_n|^2(\psi_{n+1}+\psi_{n-1})\,.
\label{final-IDNLS2}
\end{equation}
We see that the IDNLS equation is defocusing whenever 
a lattice parameter $\delta$ is smaller than 1. 
When $\delta=1$, the IDNLS equation reduces to a linear equation. 
When $\delta>1$, i.e. $a<1$, the IDNLS equation is focusing. 
In this case, above single component Casorati determinant solution 
does not satisfy the conditions for complex conjugacy
(unless we admit singular solutions). 
However, our Casorati determinant form can be extended into the case of 
the focusing IDNLS equation. 
This is corresponding to the homoclinic orbit
solution of the IDNLS equation. 
The detail will be discussed in elsewhere.   

\section{Soliton Solutions}
\begin{figure}[t!]
\centerline{
\includegraphics[scale=0.6]{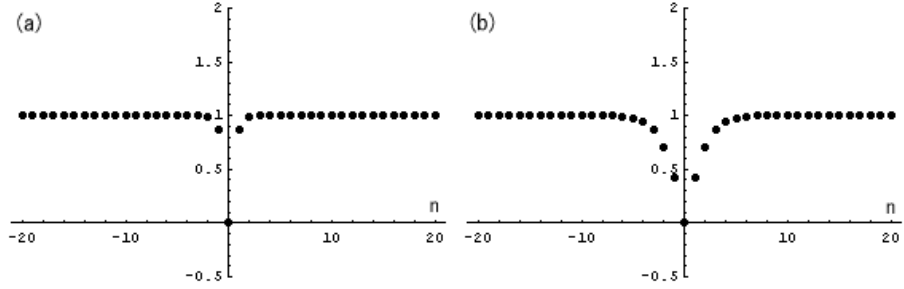}}
\caption{Plots of 1-dark soliton. The vertical axis is 
$|\psi_n|$. 
((a) $a=2,r=\exp({\rm i}\pi),\theta=\pi$,
(b) $a=1.1,r=\exp({\rm i}\pi),\theta=\pi$.)}\label{fig1}
\end{figure}

\begin{figure}[t!]
\centerline{
\includegraphics[scale=0.6]{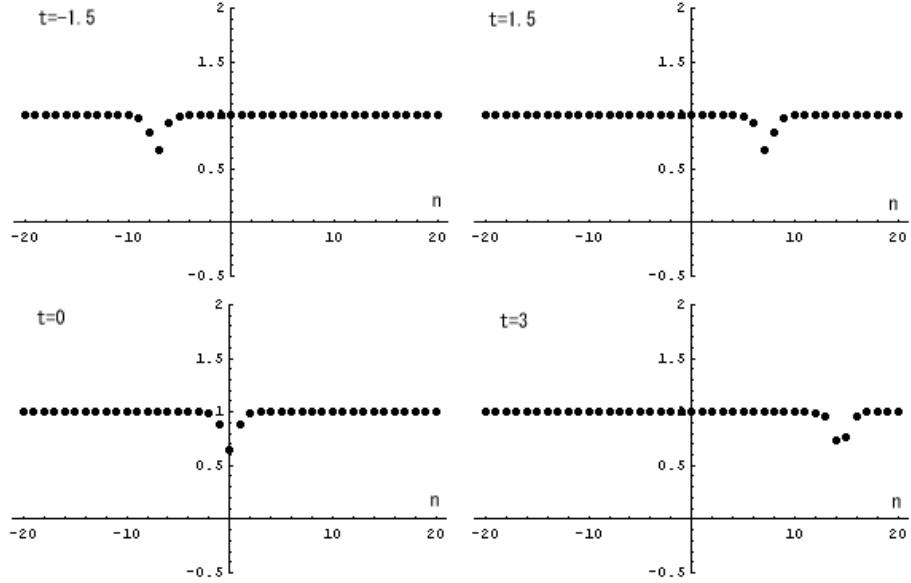}
}
\caption{Plots of travelling 1-gray soliton.
The vertical axis is 
$|\psi_n|$. 
((a) $a=2,r=\exp(3{\rm i}\pi/4),\theta=5\pi/3$.)
}\label{fig2}
\end{figure}

\begin{figure}[t!]
\centerline{
\includegraphics[scale=0.6]{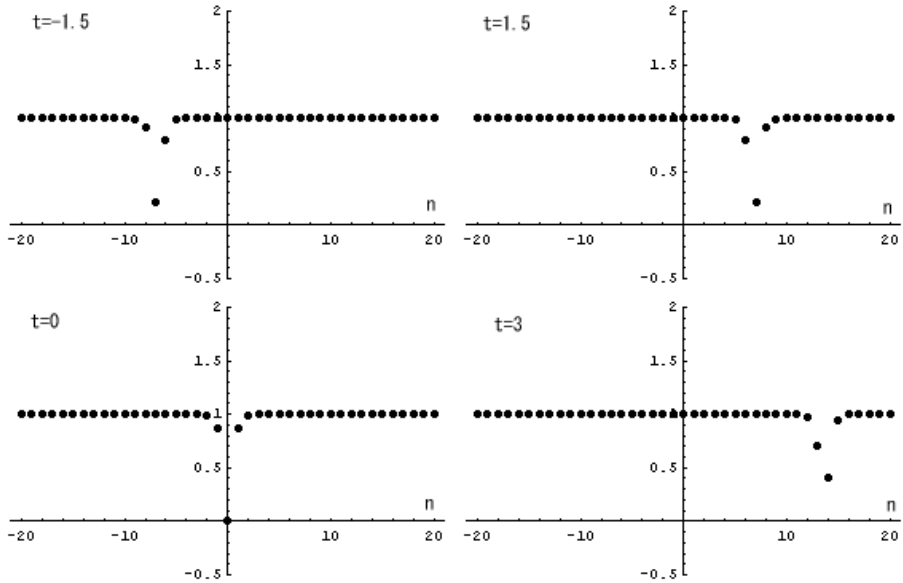}
}
\caption{Plots of travelling 1-black soliton.
The vertical axis is 
$|\psi_n|$. 
($a=2,r=\exp(-5{\rm i}\pi/3),\theta=5\pi/3$.)
}\label{fig3}
\end{figure}

In this section we give examples of dark soliton solutions
for the IDNLS equation.
By taking $N=1$ in eqs.~(\ref{expF}) and (\ref{expG}),
we get the 1-soliton solution for the IDNLS eq.~(\ref{final-IDNLS}),
$$
\psi_n={G_n \over F_n}\exp({\rm i}(\theta n-2t\cos\theta))\,,
$$
where $\theta$ is a real parameter and
$$
F_n=1+|p_1|^{2n}\exp(\xi_1+\xi_1^{*})\,,\qquad
G_n=1-{p_1 \over p_1^{*}r_1^2}|p_1|^{2n}\exp(\xi_1+\xi_1^{*})\,,
$$
where
$$
\xi_1 = {\rm i} a\left({p_1 \over \exp({\rm i}\theta)}
+{\exp({\rm i}\theta) \over p_1}\right)t + \xi_{10},\qquad
p_1=a+\sqrt{a^2-1}r_1,\qquad |r_1|=1\,.
$$
By rewriting as $p_1=p\exp({\rm i}\theta)$ and
$r_1=r\exp({\rm i}\theta)$, we get a slightly simpler expression,
$$
F_n=1+|p|^{2n}\exp(\xi+\xi^{*})\,,\qquad
G_n=1-{p \over p^{*}r^2}|p|^{2n}\exp(\xi+\xi^{*})\,,
$$
where
$$
\xi = {\rm i} a\left(p+{1 \over p}\right)t + \xi_{0},\qquad
p=a\exp(-{\rm i}\theta)+\sqrt{a^2-1}r,\qquad |r|=1\,.
$$
Here $\arg(r)$ parametrizes the wave number of soliton,
and $\Re(\xi_{0})$ gives the phase constant which parametrizes
the position of soliton. 
Figure \ref{fig1} shows a stationary 1-dark (black) soliton. 
When $a\,(>1)$ is getting closer to 1, 
the width of a dip is getting wider. 
Figures \ref{fig2} and \ref{fig3} 
show travelling 1-dark (gray and black) solitons. 

Here we should comment about the difference of travelling velocity
between black and gray solitons.
In the continuous case, the NLS equation (\ref{contiNLS}) is invariant
under the gauge and Galilean transformations,
$$
\psi(x,t)\to\psi(x-2kt,t)\exp({\rm i}(kx-k^2t)),
$$
thus the travelling velocity of any solution can be shifted by $2k$
by using the above transformation.
If we normalize the freedom of Galilean transformation by requiring
that the carrier wave disappears (i.e. $k=0$) and $\psi$ represents
only the envelope, then the velocity of the envelope soliton is $0$
if and only if the soliton is black.
In the case of IDNLS, the situation is almost same.
When $\theta=0$ (i.e. the case of carrier-wave-less) or $\theta=\pi$
(i.e. $\psi_n=(-1)^n\times\hbox{(envelope soliton)}$), then
the velocity of black soliton is $0$ and
the travelling dark solitons are always gray solitons. 
We see this fact from Figs. \ref{fig1} and \ref{fig3}. 

\begin{figure}[t!]
\centerline{
\includegraphics[scale=0.6]{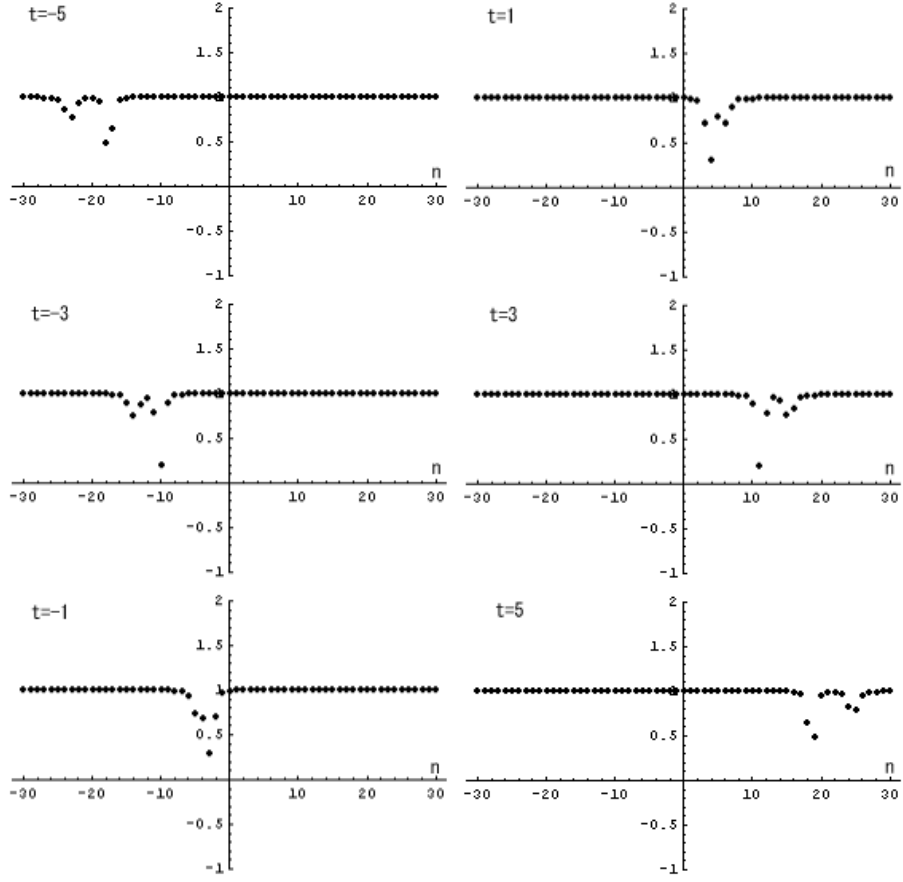}}
\caption{Plots of 2-dark soliton interaction. 
The vertical axis is $|\psi_n|$.
($a=2,r_1=\exp(3{\rm i}\pi/4), 
r_2=\exp({\rm i}\pi/4),\theta=5\pi/4$)}\label{fig4}
\end{figure}

The 2-soliton solution is obtained by taking $N=2$ in
eqs.~(\ref{expF}) and (\ref{expG}).
Similarly by rewriting as $p_i\to p_i\exp({\rm i}\theta)$ and
$r_i\to r_i\exp({\rm i}\theta)$, we get an explicit expression
of the 2-soliton solution for eq.~(\ref{final-IDNLS}),
$$
\psi_n={G_n \over F_n}\exp({\rm i}(\theta n-2t\cos\theta))\,,
$$
\begin{eqnarray*}
&&F_n=1+|p_1|^{2n}\exp(\xi_1+\xi_1^{*})
+|p_2|^{2n}\exp(\xi_2+\xi_2^{*})\\
&&+\left|{p_1-p_2 \over p_1p_2^{*}-1}\right|^2
|p_1p_2|^{2n}\exp(\xi_1+\xi_1^{*}+\xi_2+\xi_2^{*})\,,\\
&&G_n=1-{p_1 \over p_1^{*}r_1^2}|p_1|^{2n}\exp(\xi_1+\xi_1^{*})
-{p_2 \over p_2^{*}r_2^2}|p_2|^{2n}\exp({\xi_2+\xi_2^{*}})\\
&&\qquad
+\left|{p_1-p_2 \over p_1p_2^{*}-1}\right|^2
{p_1 \over p_1^{*}r_1^2}{p_2 \over p_2^{*}r_2^2}
|p_1p_2|^{2n}\exp(\xi_1+\xi_1^{*}+\xi_2+\xi_2^{*})\,,
\end{eqnarray*}
where
$$
\xi_i = {\rm i} a\left(p_i+{1 \over p_i}\right)t + \xi_{i0},\qquad
p_i=a\exp(-{\rm i}\theta)+\sqrt{a^2-1}r_i,\qquad |r_i|=1\,.
$$
Here the carrier wave of the dark soliton solution is given by
the exponential factor, $\exp({\rm i}(\theta n-2t\cos\theta))$. 
Figure \ref{fig4} shows 2-dark soliton interaction.  

%%%%%%%%%%%%%%%%%%%%%%%%%%%%%%%%%%%%%%%%%%%%%%%%%%%%%%%%%%%%%%%%%%%%%%%%%

\section{Concluding Remarks}

We have shown that the $N$-dark soliton solutions of 
the IDNLS equation are given by the Casorati determinant.
{}From the above derivation of the $N$-dark soliton solutions 
for the IDNLS equation, we notice that there is 
correspondence between the Casorati 
determinant solutions of the IDNLS equation and 
the RTL equation \cite{OKMS}. 
The RTL equation is decomposed into three bilinear equations, 
i.e. the Toda lattice (TL) equation, the B\"acklund
transformation (BT) for TL and the DTL equation 
(See appendix A \ref{appRTL}). 
The IDNLS equation is also decomposed into three bilinear 
equations, i.e. two types of the BT for TL 
and the DTL equation.  

This fact reminds us several works 
on the relationship between IDNLS and RTL 
\cite{Vekslerchik,KMZ97,Sadakane}. 
Moreover, our result is an answer to an open problem of 
ref.\citen{Sadakane}, i.e., 
the $N$-bright soliton solutions of IDNLS are written by 
the 2-component Casorati determinant in ref.\citen{Sadakane}, 
and the $N$-dark soliton solution are written by 
the single-component Casorati determinant 
which is given in this paper and corresponds to 
the solution of RTL in ref.\citen{OKMS}. 

It is interesting to investigate $N$-dark soliton solutions 
of the discrete time IDNLS equation.   
It will be reported in the forthcoming paper. 

\vskip 2mm
\section*{Acknowledgment}
Prof. M. J. Ablowitz, Prof. Y. Kodama, Dr. T. Tsuchida, Prof. B. F. Feng, 
Prof. K. Kajiwara and Dr. A. Mukaihira. 
K.M.\ acknowledges support from the 21st Century 
COE program ``Development of Dynamic Mathematics with High Functionality'' 
at Faculty of Mathematics, Kyushu University.

\appendix
\section{Relativistic Toda Lattice}\label{appRTL}
In this appendix we briefly explain $\tau$-functions of 
the RTL equation.
The RTL equation
\begin{eqnarray}
\frac{d^2 q_n}{dt^2} 
&=&( 1 + {1 \over {\rm c}}\dot q_{n-1})
( 1 + {1 \over {\rm c}}\dot q_n)
{\exp(q_{n-1}-q_n) \over
1+{1 \over {\rm c}^2}\exp(q_{n-1}-q_n)}\nonumber\\
&-&(1+{1 \over {\rm c}}\dot q_n)
( 1 + {1 \over {\rm c}}\dot q_{n+1})
{\exp(q_n-q_{n+1}) \over
1+{1 \over {\rm c}^2}\exp(q_n-q_{n+1})},\label{rtl}
\end{eqnarray}
where $q_n$ is the coordinates of $n$-th lattice point 
and ${\rm c}$ is the light
speed, was introduced and studied by Ruijsenaars \cite{RTL}. 
The RTL equation (\ref{rtl}) 
is transformed into the three bilinear equations,
\begin{eqnarray}
&&D_x^2f_n\cdot f_n=2({\bar g}_ng_n - f_n^2 ),\nonumber\\
&&(aD_x-1)f_n\cdot f_{n-1}+{\bar g}_{n-1}g_n = 0,\label{bi-rtl}\\
&&{\bar g}_{n-1}g_{n+1}-f_n^2= 
a^2(f_{n+1}f_{n-1}-f_n^2 ),\nonumber
\end{eqnarray}
through the variable transformations,
\begin{eqnarray}
&&q_n = \log{f_{n-1} \over f_n},\\
&&t = {\sqrt{1+{\rm c}^2} \over {\rm c}}x,\\
&&{\rm c} = {\sqrt{1-a^2} \over a}.
\end{eqnarray}
In eqs.~(\ref{bi-rtl}), 
$g$ and $\bar g$ are the auxiliary variables and $D$ is
the bilinear differential operator defined by
$$
 D_x^k f\cdot g = (\partial_x-\partial_y)^k f(x)g(y)|_{y=x}.
$$

Letting
\begin{eqnarray}
f_n = \tau_n(n),\quad g_n = \tau_{n-1}(n),
\quad {\bar g}_n = \tau_{n+1}(n),\label{rtl-trans}
\end{eqnarray}
we have three bilinear equations, 
the TL equation, the B\"acklund
transformation (BT) for TL and the DTL equation,
\begin{eqnarray}
&& D_x^2\tau_n(k)\cdot\tau_n(k)
=2(\tau_{n+1}(k)\tau_{n-1}(k)-\tau_n(k)\tau_n(k)),\nonumber\\
&&(aD_x-1)\tau_n(k)\cdot\tau_{n-1}(k-1) + \tau_n(k-1)\tau_{n-1}(k)
 = 0,\\
&&\tau_n(k+1)\tau_n(k-1)-\tau_n(k)\tau_n(k)\nonumber\\
&&\quad \quad 
=a^2(\tau_{n+1}(k+1)\tau_{n-1}(k-1)-\tau_n(k)\tau_n(k)).\nonumber
\end{eqnarray}

\section{Proof of Eqs. (\ref{expF})-(\ref{expH})}\label{appproof}
Let us first consider the Casorati determinant
$\displaystyle D=\det_{1\le i,j\le N}(p_i^{j-1}A_i+q_i^{j-1}B_i)$.
Since each row is sum of two vectors, the determinant is given by
the sum of $2^N$ terms,
\begin{eqnarray*}
&&D=\left|\begin{matrix}
B_1 &q_1B_1 &\cdots &q_1^{N-1}B_1 \cr
B_2 &q_2B_2 &\cdots &q_2^{N-1}B_2 \cr
\vdots &\vdots &&\vdots \cr
B_N &q_NB_N &\cdots &q_N^{N-1}B_N \cr
\end{matrix}\right|
+\sum_{i_1=1}^N\left|\begin{matrix}
B_1 &q_1B_1 &\cdots &q_1^{N-1}B_1 \cr
\vdots &\vdots &&\vdots \cr
A_{i_1} &p_{i_1}A_{i_1} &\cdots &p_{i_1}^{N-1}A_{i_1} \cr
\vdots &\vdots &&\vdots \cr
B_N &q_NB_N &\cdots &q_N^{N-1}B_N \cr
\end{matrix}\right|
\\
&&+\sum_{1\le i_1<i_2\le N}\left|\begin{matrix}
B_1 &q_1B_1 &\cdots &q_1^{N-1}B_1 \cr
\vdots &\vdots &&\vdots \cr
A_{i_1} &p_{i_1}A_{i_1} &\cdots &p_{i_1}^{N-1}A_{i_1} \cr
\vdots &\vdots &&\vdots \cr
A_{i_2} &p_{i_2}A_{i_2} &\cdots &p_{i_2}^{N-1}A_{i_2} \cr
\vdots &\vdots &&\vdots \cr
B_N &q_NB_N &\cdots &q_N^{N-1}B_N \cr
\end{matrix}\right|
+\cdots+\left|\begin{matrix}
A_1 &p_1A_1 &\cdots &p_1^{N-1}A_1 \cr
A_2 &p_2A_2 &\cdots &p_2^{N-1}A_2 \cr
\vdots &\vdots &&\vdots \cr
A_N &p_NA_N &\cdots &p_N^{N-1}A_N \cr
\end{matrix}\right|
\\
&&=\Delta(q_1,q_2,\cdots,q_N)\prod_{i=1}^NB_i
+\sum_{i_1=1}^N\Delta(q_1,\cdots,p_{i_1},\cdots,q_N)
A_{i_1}\prod_{{i=1 \atop i\ne i_1}}^NB_i
\\
&&+\sum_{1\le i_1<i_2\le N}
\Delta(q_1,\cdots,p_{i_1},\cdots,p_{i_2},\cdots,q_N)
A_{i_1}A_{i_2}\prod_{{i=1 \atop i\ne i_1,i_2}}^NB_i+\cdots
\\
&&+\sum_{1\le i_1<i_2<\cdots<i_M\le N}
\Delta(q_1,\cdots,p_{i_1},\cdots,p_{i_2},\cdots,p_{i_M},\cdots,q_N)
A_{i_1}A_{i_2}\cdots A_{i_M}
\prod_{{i=1 \atop i\ne i_1,i_2,\cdots,i_M}}^NB_i
\\
&&+\cdots+\Delta(p_1,p_2,\cdots,p_N)\prod_{i=1}^NA_i
\\
&&=\Delta(q_1,q_2,\cdots,q_N)\prod_{i=1}^NB_i\left(1+\sum_{i_1=1}^N
\frac{\Delta(q_1,\cdots,p_{i_1},\cdots,q_N)}{\Delta(q_1,\cdots,q_N)}
\frac{A_{i_1}}{B_{i_1}}\right.
\\
&&+\sum_{1\le i_1<i_2\le N}
\frac{\Delta(q_1,\cdots,p_{i_1},\cdots,p_{i_2},\cdots,q_N)}
{\Delta(q_1,\cdots,q_N)}\frac{A_{i_1}A_{i_2}}{B_{i_1}B_{i_2}}+\cdots
\\
&&+\sum_{1\le i_1<i_2<\cdots<i_M\le N}
\frac{\Delta(q_1,\cdots,p_{i_1},\cdots,p_{i_2},\cdots,p_{i_M},\cdots,q_N)}
{\Delta(q_1,\cdots,q_N)}\prod_{\nu=1}^M\frac{A_{i_\nu}}{B_{i_\nu}}
\\
&&\left.+\cdots+\frac{\Delta(p_1,\cdots,p_N)}{\Delta(q_1,\cdots,q_N)}
\prod_{i=1}^N\frac{A_i}{B_i}\right),
\end{eqnarray*}
where $\Delta$ is the Vandermonde determinant defined by
$$
\Delta(x_1,x_2,\cdots,x_N)=\prod_{N\ge i>j\ge1}(x_i-x_j).
$$
By rewriting as
$$
B_i=\left(\prod_{{k=1 \atop k\ne i}}^N
{p_i-q_k \over q_i-q_k}\right)B'_i,
$$
each term of the summation is given by
\begin{eqnarray*}
&&\frac
{\Delta(q_1,\cdots,p_{i_1},\cdots,p_{i_2},\cdots,p_{i_M},\cdots,q_N)}
{\Delta(q_1,\cdots,q_N)}\prod_{\nu=1}^M\frac{A_{i_\nu}}{B_{i_\nu}}
\\
&&=\frac{\Delta(p_{i_1},p_{i_2},\cdots,p_{i_M})}
{\Delta(q_{i_1},q_{i_2},\cdots,q_{i_M})}
\left(\prod_{\nu=1}^M\prod_{{j=1 \atop j\ne i_1,i_2,\cdots,i_M}}^N
{p_{i_\nu}-q_j \over q_{i_\nu}-q_j}\right)
\prod_{\nu=1}^M\left(\prod_{{k=1 \atop k\ne {i_\nu}}}^N
{q_{i_\nu}-q_k \over p_{i_\nu}-q_k}\right)\frac{A_{i_\nu}}{B'_{i_\nu}}
\\
&&=\frac{\Delta(p_{i_1},p_{i_2},\cdots,p_{i_M})}
{\Delta(q_{i_1},q_{i_2},\cdots,q_{i_M})}
\left(\prod_{\nu=1}^M\prod_{{\mu=1 \atop \mu\ne\nu}}^M
{q_{i_\nu}-q_{i_\mu} \over p_{i_\nu}-q_{i_\mu}}\right)
\prod_{\nu=1}^M\frac{A_{i_\nu}}{B'_{i_\nu}}
=\left(\prod_{1\le\mu<\nu\le M}
{(p_{i_\mu}-p_{i_\nu})(q_{i_\mu}-q_{i_\nu})
\over (p_{i_\mu}-q_{i_\nu})(q_{i_\mu}-p_{i_\nu})}\right)
\prod_{\nu=1}^M\frac{A_{i_\nu}}{B'_{i_\nu}},
\end{eqnarray*}
thus we obtain
$$
D=\left(\Delta(q_1,q_2,\cdots,q_N)\prod_{i=1}^NB_i\right)
\sum_{M=0}^N\ \sum_{1\le i_1<i_2<\cdots<i_M\le N}
\left(\prod_{1\le\mu<\nu\le M}
{(p_{i_\mu}-p_{i_\nu})(q_{i_\mu}-q_{i_\nu})
\over (p_{i_\mu}-q_{i_\nu})(q_{i_\mu}-p_{i_\nu})}\right)
\prod_{\nu=1}^M\frac{A_{i_\nu}}{B'_{i_\nu}}.
$$
The prefactor of above summation gives the gauge factor
in eq. (\ref{gaugetrans}), and from the summation part,
$F_n$, $G_n$ and $H_n$ in eqs. (\ref{expF})-(\ref{expH})
are derived by taking
$$
q_i={1 \over p_i^{*}},\qquad p_i=a+\sqrt{a^2-1}r_i,\qquad |r_i|=1,
$$
and
$$
\begin{array}{lll}
A_i=p_i^n\exp(\xi_i),
&\displaystyle B'_i={1 \over {p_i^{*}}^n}\exp(-\xi_i^{*}),
&\hbox{for }F_n,
\\[10pt]
\displaystyle A_i={p_i^{n+1} \over 1-ap_i}\exp(\xi_i),
&\displaystyle
B'_i={1 \over {p_i^{*}}^{n+1}(1-a/p_i^{*})}\exp(-\xi_i^{*}),
&\hbox{for }G_n,
\\[10pt]
A_i=p_i^{n-1}(1-ap_i)\exp(\xi_i),
&\displaystyle
B'_i={1-a/p_i^{*} \over {p_i^{*}}^{n-1}}\exp(-\xi_i^{*}),
&\hbox{for }H_n.
\end{array}
$$

\end{document}